\documentclass{article}

\usepackage{arxiv}

\usepackage[utf8]{inputenc} 
\usepackage[T1]{fontenc}    
\usepackage{hyperref}       
\usepackage{url}            
\usepackage{booktabs}       
\usepackage{amsfonts}       
\usepackage{nicefrac}       
\usepackage{microtype}      
\usepackage{lipsum}
\usepackage{graphicx}
\graphicspath{ {./images/} }

\title{How smart is the grid?}

\author{
Ermanno Lo Cascio* \\
Industrial Process and Energy Systems
Engineering (IPESE) Group\\ 
École Polytechnique Fédérale de Lausanne\\
EPFL Valais-Wallis, CH-1951 Sion\\
\texttt{Ermanno.locascio@epfl.ch}  
  \And
   Zhenjun Ma \\
   Sustainable Buildings Research Centre (SBRC) \\
   University of Wollongong \\
   Wollongong - 2522, Australia \\
   \texttt{zhenjun@uow.edu.au} \\
  \And
   François Maréchal \\
   Industrial Process and Energy Systems Engineering (IPESE) Group \\
   École Polytechnique Fédérale de Lausanne\\ 
   EPFL Valais-Wallis, CH-1951 Sion \\
  \texttt{francois.marechal@epfl.ch} \\
}

\begin{document}
\maketitle
\begin{abstract}
Ancient Romans called \emph{urbs} the set of buildings and
infrastructures, and \emph{civitas} the Roman citizens\emph{.} Today
instead, while the society is surfing the digital tsunami, \emph{urbs}
and \emph{civitas} tend to become much closer, almost merging, that we
might attempt to condensate these into a single concept: \emph{smart
grid}. Internet of things, artificial intelligence, blockchain, quantum
cryptography is only a few of the paradigms that are likely to
contribute to determining the final portrait of the future smart grid. However, to understand the effective sustainability of complex grids, specific tools are required. To this end, in this article, a systematic review of the emerging paradigms is
presented, identifying intersectoral synergies and limitations with
respect to the `smart grid' concept. Further, a taxonomic framework for assessing the level of sustainability of the grid is proposed. Finally, from the scenario
portrayed, a set of issues involving engineering, regulation, security,
and social frameworks have been derived in a theoretical fashion.
The findings are likely to suggest the urgent need for multidisciplinary
cooperation to wisely address engineering and ontological challenges
gravitating around the smart grid concept.
\end{abstract}

\emph{Highlights }

\begin{itemize}
\item
  A systematic holistic review of the smart grid emerging paradigms is
  presented.
\item
  The blurry dichotomy `synergies vs
  complexity' deriving from integration of technological domains is theoretically explored.~
\item
  A taxonomic framework for smart grid categorization is proposed.
\item
  System complexity might be the Achille's heel of the emerging
  monumental grid.~
\item 
 The urgent need of exploring faults propagation mechanisms
  in complex interdependent networks is identfied.~~
\item
 Privacy, e-addiction, and cyber security issues need to be regulated
  in a resolutive fashion in the short term, while space programs might be needed in the long term to ensure space security and research on Sun cycles.~
\end{itemize}

\emph{Keywords}: cascading failures; Internet of Things; resilience; domino effect;
smart grid generations; cyber-crime; integrated energy systems; planetary grids; World energy transition

* Author to whom correspondence should be addressed. E-Mail:
ermanno.locascio@epfl.ch


\section{Introduction }\label{introduction}

With wisdom and paternal calm, our globe is warmly hosting a massive
revolution based on enormous immaterial fluxes of information. On the
stage of the 4.0 theatre, it seems that we are subject and object at the
same time: the huge digital carousel it's being constantly spin by
commercial paradoxes where people are customers of their product: data.
Whether it is ``good'' or ``bad'', well, the show must go on. In
fact, during the past two years, the 90 \% of the data in the world were
created and 2.5 quintillion bytes of data are created every day \cite{FAO2019Data_Production}.
This is thanks to the digital technologies that have also made expand
the sectorial conceptual borders, especially for the smart grid
archetype, where end-users and complementary sectors like
transportation, tends to be intimately linked. This is also thanks to
the advances in computing power and efficiency have enabled more
powerful and sophisticated analytic, such as artificial intelligence
and automation \cite{Digitalization}. According to the International Energy Agency \cite{IEA}, digital technologies can help make the energy system more
intelligent, reliable and sustainable, whereas it is also raising
security and privacy risks, changing market.' However, if we put
ourselves in a meta-perspective and, if we reframe this scenario, we
might also convince that the market is changing the
digitalization, making the energy system more connected for sure. But,
intelligent? Resilient? Sustainable? The exuberant availability of
electronic devices \cite{sovacool2020smart}, for instance, seems to be the proof of the presence
of an uncontrolled commercial speculative pool whose inertia, if not
properly addressed, would likely affect the evolution of the smart grid,
exchanging threats with strengths.
Thus, from this scenario, it emerges the desire of attempting answering the question: ‘how smart is the grid?’. However, to this aim, first we need to delineate what the ‘grid’ refers to, and what ‘smart’ stands for. Factually, from today’s perspective, the term ‘grid’ assumes a broader meaning involving not only the electrical transmission network, but also other energy carriers (thermal, natural gas, etc.) as they have become a fundamental part of the grid, and their operations, services delivery, and management tend to be highly interdependent, e.g. smart thermal grids \cite{lund20144th} and gas grids \cite{lund2017smart}. 

On the other hand, the term ‘smart’ suffers of a high semantic inflation, which commonly pushes us to exchange the ‘smartness’ of the ‘grid’ with its level of automation, while the level of automation might not always be synonymous of smartness - in the most general sense of the term. Thus, in this study, we associate the term ‘smart’ to the final objective of \textit{increasing the life-quality of the community by creating a holistically sustainable energy system and service}. Given this premise, at the state-of-the-art, different scientists have provided structured discussions for the topic. Some authors contributed by reviewing the issue through a vertical approach, thus providing an in-depth review of a given technological sub-domain of the smart grid framework. In fact, from one hand, for instance, Tu et al. \cite{tu2017big} reviewed the big data issues for the smart grid, and thoroughly discussed theoretical and practical applications, with reference to the power grid. In \cite{sovacool2020smart}, the authors examined the technologies for smart homes in Europe, and discussed concepts, benefits, risks and policies. In \cite{reka2018future} and in \cite{stojkoska2017review}, the most significant research studies on the application of the Internet of Things technologies for the smart grid framework were reviewed. On the other hand, instead, some authors reviewed the problem from a broader perspective. Dileep \cite{dileep2020survey}, for instance, provided an extensive survey of the domain, and thoroughly discussed technologies and applications, with a particular focus on the electrical grid. Similarly, Tuballa and  Abundo \cite{tuballa2016review} presented an overview of the smart grid, and discussed its features, functionalities and characteristics. Birbi and Krogstie \cite{bibri2017smart} provided a comprehensive overview of the domain evaluating the foundations and assumptions of the smart (and) sustainable cities. Here, the authors identified the need to develop a theoretical and practically convincing framework for strategic sustainable urban development. 
The level of complexity of the smart grid domain, specially due to the ICT penetration, make it difficult to understand the effective sustainability of the technological configuration adopted. For this reason, the authors identify the need to outline a preliminary assessment framework. To this aim,  
in the presented paper, a new reviewing approach has been employed. In fact, while these last studies provide a review whose conceptual borders involve a given domain and/or a given issue of the framework, in this article, we attempt to answer to ontological questions by employing a broader analytical approach, using a multi-domain anthology as a tool, thus embracing those paradigms, issues and aspects that are likely to come into play in the definition of the portrait of the future smart grid. As outcome of this review process, we obtain and propose a preliminary socio-technological taxonomic model which is likely to begin the definition of a rigorous labelling protocol to assess - in a objective way - the smartness of a generic grid. Finally, we aim to identify possible mid and long-term actions that
hopefully help make us realize a truly sustainable urban future.

\section{Methodology}\label{methodology}

Fig. 1 represents the methodology employed in this research to outline a taxonomic model to assess the sustainability of the grid, and to derive mid and long term actions. To this end, a complete collection of paradigms is constructed (smart grid anthology). This last is employed to drive a higher level process, where a systematic and critical analysis is conducted, working on the blurry line between synergies and complexity emerging from novel paradigms and/or domains inter-dependencies. Finally, starting from the new portrait of the smart grid, we attempt to explicitly re-frame the fundamental and implicit values of the smart grid concept (ontology). 

\begin{figure}
\begin{center}
    \includegraphics[width=1\textwidth]{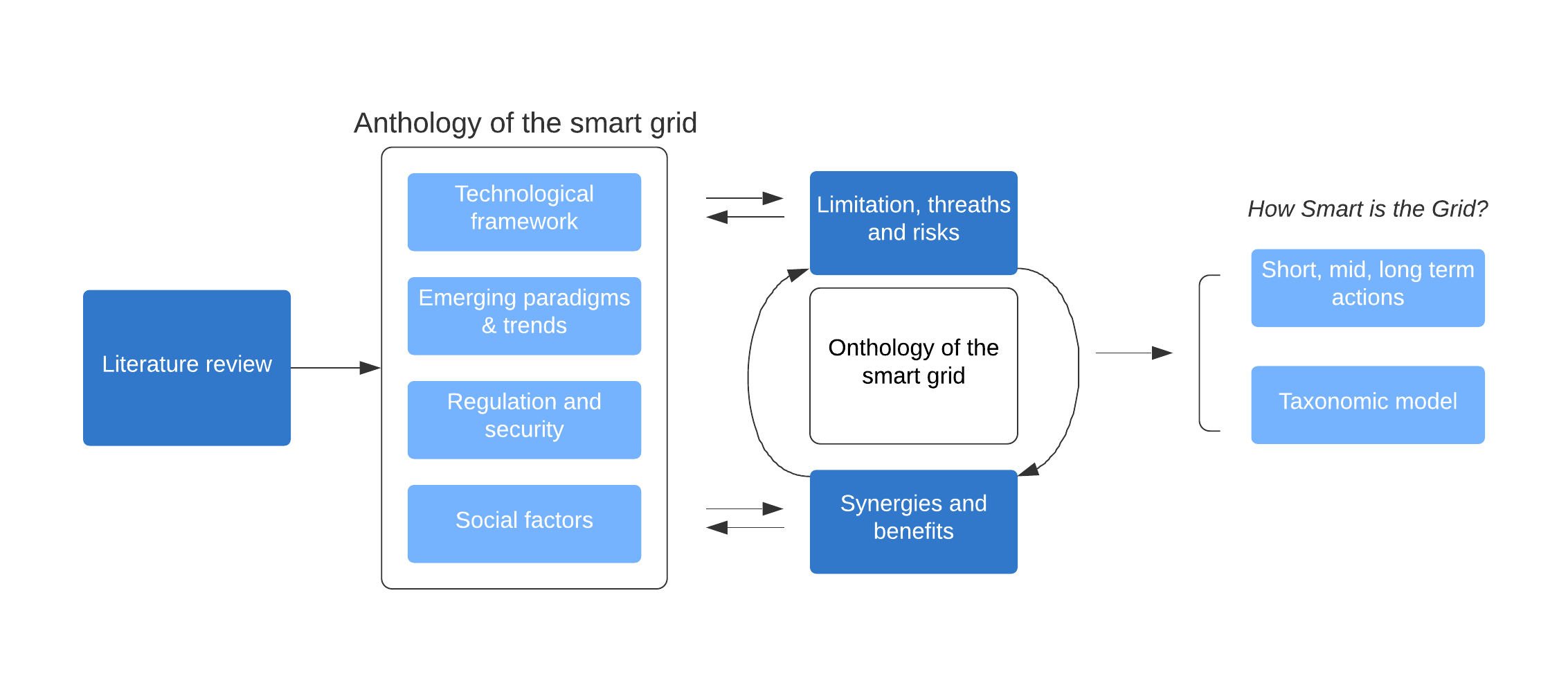}
\end{center}
\caption{Study methodology.}
\end{figure}

\section{Review: technological framework, emerging trends and
  paradigms}\label{technological-framework-emerging-trends-and-paradigms}

\subsection{Internet of Things }\label{internet-of-things}

The~Internet of Things~(IoT) is the network of devices that have network
connectivity and sensors to acquire information from the external
environment. Sofana et al. \cite{reka2018future} provided a review of the ongoing
research contextualizing -- in a comprehensive fashion -- the role of
IoT within the smart grid framework. This research outlines a
technological perspective according to which the IoT evolution needs
further developments at architectural and standardization levels.
Similarly, Zhou et al. \cite{zhou2016smart} provided an overview of the smart home
energy management systems, highlighting the need for further conceptual,
technological and architectural upgrades to welcome and better exploit
renewable energy technologies in the residential and tertiary sector. In
\cite{schieweck2018smart}, Schieweck et al. analyzed the e-panorama for smart homes
focusing on indoor air quality and people's perception. From the context
pictured by this review study, it clearly emerges that there is a great
margin of improvement at different levels, especially referred to the
human adaptation and interaction with the smart home environment. But,
do IoTs technologies could help reduce energy consumption by means
of optimal control of the heating, ventilation, and air conditioning
(HVAC) system? It depends from the context considered and the "point of view". In fact, if we look at the building level, for instance, Lomas et al. \cite{lomas2018domestic} recognized moderated
quality evidence that smart thermostats may not save energy compared to
a non-smart thermostat. Even, for a case study in the USA, the energy
demand increased by 2-4\% with self-learning algorithms, compared with
conventional on-off control. However, if we look at the system level (generation, distribution, end-users) the effective energy impact originated by the use of smart thermostats or, in general, by in-home display for HVAC control, might be sensibly different. In fact, smart thermostat can participate in utility demand response programs \cite{king2018energy}, thanks to the connectivity enabled among the system's players, thus representing a strategic element of the "puzzle" to unlock sustainable operations, without affecting the users' comfort, through peak shaving and demand-side management, specially when aggregated across different homes \cite{Energy_News}. Furthermore, Wang \cite{wang2018transactive} glimpsed a
great opportunity behind smart in-home devices if these are capable to
transact energy for peer-to-peer applications or with the grid.
Generally speaking, it is likely to be that every case, every building,
has its own characteristics in terms of energy-saving potential and
controllability which are mainly linked to the boundary conditions
(weather, etc.), the envelope thermophysical characteristics, HVAC type
and control and, finally, occupancy profiles and human factors. The
peculiarities of every building are probably the reason that brings
Balta-Ozkan \cite{balta2013development} to identify a need for a holistic view for the
design and delivery of smart home services, enabling tailored solutions
for householders, appropriate to the context. As shown in figure 2, the IoT global market for
end-users is expected to grow up to 1.6 trillion in US dollars by 2025
\cite{IoT_market}. According to \emph{Fortune Business Inside} \cite{IoT_market_investment} instead,
the IoT market is expected to reach 1.1 trillion US dollars in 2026. In
any case, the sophisticated e-cosmo is actually a multidomain connected,
fast-interacting set of physical players (subjects and objects) and
every measurable evolution, even its associable economic growth, will be
certainly related by this existing sectorial interdependency. In this
perspective, an example of technology-to-technology synergy could be
represented by the paradigm called blockchain, which is likely to be a
game-changer tool for peer-to-peer energy transactions while it will
work as a catalyzer for the IoT market growth. But, the `evolution
equation' of the smart grid is constituted by further several variables
that will determine the final picture of the digital era e.g. innovation
in telecommunication, information technologies, regulation and as well
as anthropological issues. In the following section, we propose an
in-depth analysis of those further main archetypes involved.

\begin{figure}
\begin{center}
    \includegraphics[width=0.6\textwidth]{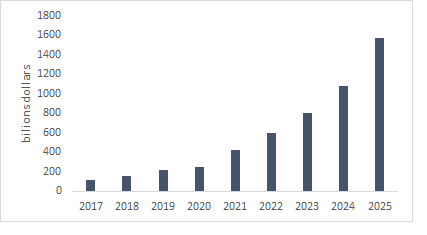}
\end{center}
\caption{Forecast end-user spending on IoT solutions worldwide from 2017 to 2025. Derived from \cite{IoT_market}.}
\end{figure}

\subsection{Smart meters }\label{smart-meters}

A meter is expected to have the following capability to be categorized
as a `smart meter' \cite{alahakoon2015smart}:

\begin{itemize}
\item
  \begin{quote}
  real-time monitoring;
  \end{quote}
\item
  \begin{quote}
  remote and local data accessibility;
  \end{quote}
\item
  \begin{quote}
  remote controllability of the meter, for energy cut off;
  \end{quote}
\item
  \begin{quote}
  interaction with other meters (gas, water);
  \end{quote}
\item
  \begin{quote}
  power quality monitoring and self-analysis of the device itself;
  \end{quote}
\item
  \begin{quote}
  interaction with IoTs.
  \end{quote}
\end{itemize}

In \cite{bastida2019exploring}, Bastida et al. analyzed the energy-saving linked to the
electricity sector that could be achieved by employing ICTs in European
households. Quantitatively speaking, the contribution of ICTs would
range between 0.23\% and 3.3\% of the European CO2 reduction target with
respect to the 1.5 Celsius degrees minimal realistic warming goal at
2100. Smart metering has the potential to revolutionize access to energy
consumption data but, as highlighted by Webborn and Oreszczyn in
\cite{webborn2019champion}, a coordinated effort is needed between legislation, funding
bodies and researchers to unlock its potential. From this perspective,
the European Union issued Directives 2009/72/EC, 2009/73/EC, and
Directive 2012/27/EC that insisted on making smart meters available to the
majority of households in the EU by 2020. Italy was the first European
country where smart meters rollout started at a large scale, followed by
France who started the process in 2013, while in the UK and The
Netherlands smart meters have been introduced simultaneously in gas and
electricity sectors \cite{Rollout}. Of course, a rollout phase is a complex
process, and operators have to deal with different aspects and issues
ranging from logistic to complex aspects relative to the social domain,
passing through financial and technical challenges. In this sense, the
diffusion of a given technology is also intimately linked/bounded by the
perception and awareness of people of the technology itself. For
example, in \cite{chawla2019public}, the authors investigated the awareness and
acceptance level of smart meters among social media users in Poland.
Findings suggested a low level of public awareness for this
technology for this geographical context, thus limiting the potential
benefits that smart meters could generate for them. But, making few steps
backward, apart from the numerous sectorial difficulties, smart meters
are an essential element for the sophisticated technological
interweaving which will constitute the smart grid and its interaction
with end-users through a smart environment of the Internet of Things.

\subsection{Blockchain}\label{blockchain}

Bitcoin, based on blockchain technology, is a cryptocurrency, initially
introduced in 2009 by an author whose pseudonym is Nakamoto \cite{nakamoto2019bitcoin} and
it consists of a verification mechanism, based on distributed consensus
and cryptographic security measures. However, as previously mentioned,
blockchain is a technology that can be successfully employed also in
energy contexts. In fact, when combined with smart contracts \cite{zheng2018blockchain},
blockchain is capable to enable a decentralized market \cite{yuan2016towards}. This
aspect opens the possibility to realize what has been defined by some
scientists as the `energy democratization' where market dynamics are
induced by the community of end-users rather than a centralized
organization (figure 3). In \cite{mengelkamp2018blockchain}, Mengelkamp et al. faced the design aspect of
a local decentralized energy market based on blockchain technology. To
this aim, the authors realized a proof-of-concept model, including a
simulation of a local blockchain-based market where users can
bilaterally exchange energy. In \cite{andoni2019blockchain}, Andoni et al., based on the
review of 140 blockchain research projects, outlined the key challenges
and future outlook for this technology and its application in the energy
sector. According to the authors, blockchain represents a promising
technology -- for different sectors -- but several questions need to be
addressed in terms of technology scalability, speed, and security.
Furthermore, large consensus algorithms need to be further investigated,
from a different perspective, with attention to energy consumption and
cyber-attacks resiliency. In concluding, from this study, it emerges that,
even if they have successfully passed the proof-of-concept phase, most
projects are still in the early development stage, and thus, further
research efforts will have to demonstrate if the technology can reach
its technical viability and commercial potential \cite{andoni2019blockchain}. In this
perspective, quantum communication will likely help to make a huge step
forward in data security thanks to quantum key distribution \cite{tariq2019speculative},
that ~involves sending encrypted data as classical bits over networks,
while the keys to decrypt the information are encoded and transmitted in
a quantum state using qubits, thus enabling, in theory, an ultra-secure
communication.

\begin{figure}
\begin{center}
    \includegraphics[width=0.9\textwidth]{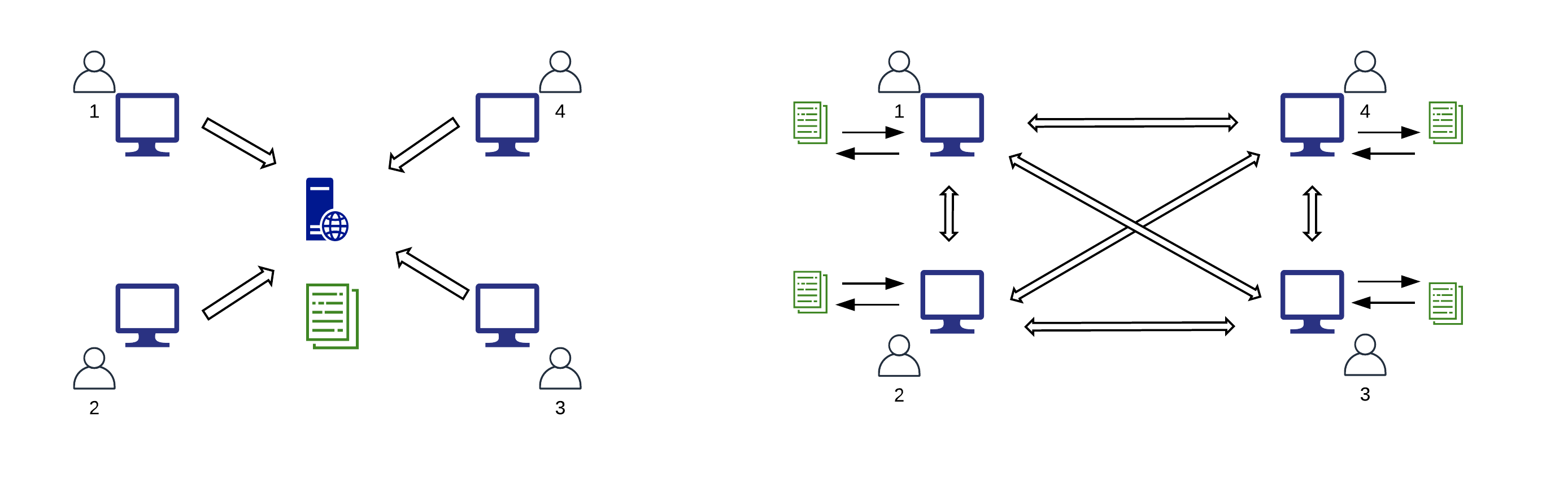}
\end{center}
\caption{Market structure: current vs blockchain-based. Derived from \cite{PwC}.}
\end{figure}

\subsection{5G/6G}\label{g6g}

5G is the fifth-generation wireless technology, whose technological
improvements, as defined by the Next \emph{Generation Mobile Network}
\emph{Alliance} \cite{osseiran2014scenarios}, involve connection speed and capacity, while
increasing the latency i.e. the time taken by devices to respond to each
other over the wireless network, thus passing from roughly 30 ms of the
4G to 1 ms. Likely, within the first years of 20', 5G will start being
deployed at a wider level and this little step will affect the society
by enabling the connection of billions of devices, affecting every
sector, e.g. health, school, and communities, thus increasing the
``smartness'' of our cities. On the other hand, industries and
businesses will be able to gather an enormous amount of information,
allowing them to achieve a level of insights capability without
precedents. In this perspective, 5G will catalyze the emergence of new
technologies such as virtual reality, or services that we cannot even
imagine from today's perspective. But, if 5G will enable communication
with unprecedented performance, on the other hand, 6G will drastically
shape the communication framework, generating new societal paradigms,
thus opening the way to new services such as holographic communication,
high precision manufacturing, allowing artificial intelligence achieving
its maximum potential \cite{strinati20196g}. From a smart city perspective instead,
according to Tariq et al. \cite{tariq2019speculative}, with 5G technology energy systems
and transportation networks are individually smart. The difference with
6G is that the control and optimization of energy and transport
infrastructure will occur in a holistic and integrated fashion, thus,
enabling a truly smart city. In \cite{saad2019vision}, the authors outlined a set of
possible perspective scenarios for 6G, identifying the complementary
technologies and infrastructure that will be likely needed. In this
sense, for instance, 6G will need an integrated terrestrial, airbone and
satellite communication network \cite{cao2018airborne}. Here, as shown in figure 4,
drones will be needed to provide connectivity to those zones where other
infrastructures are not sufficiently developed. Furthermore, both drones
and terrestrial stations may need connectivity to low orbit satellites
and CubSat \cite{cubesat}. As a summary, challenges for the 5G and 6G do not
only involve the technological side. 5G and 6G will have to be analyzed
and discussed also from a health perspective. Precisely, as reported in
\cite{di2018towards}, there is an urgent need to undertake further experimental and
epidemiologic studies to understand the effect on humans of exposure to
these specific radio frequencies.

\begin{figure}
\begin{center}
    \includegraphics[width=0.6\textwidth]{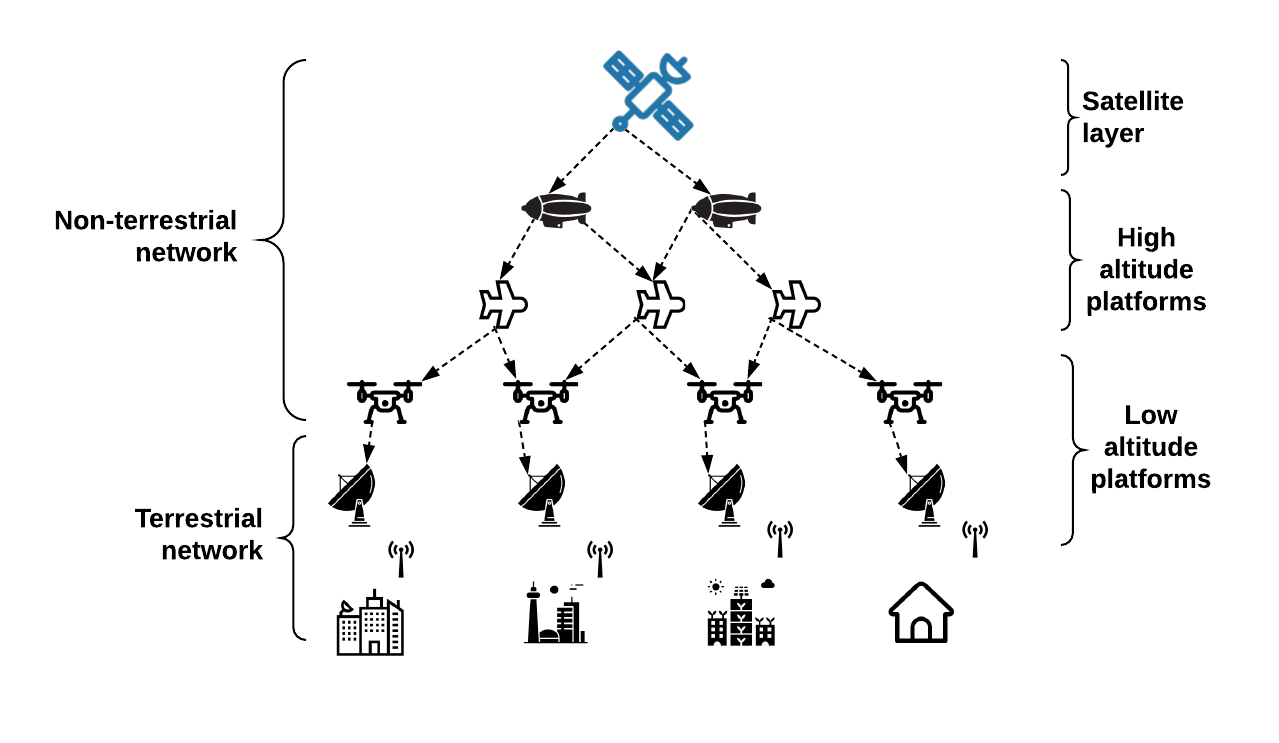}
\end{center}
\caption{An architecture airborne communication network \cite{cao2018airborne}.}
\end{figure}

\subsection{Quantum computing \& smart
grid}\label{quantum-computing-smart-grid}

Synergies emerging from energy systems integration are the main
objective pursued by most scientists of the domain to unlock the
possibility to achieve performing operations with lower environmental
impact, thus increasing the renewable energy share. Energy systems
integration and flexibility directly bring to the possibility to
optimize the energy system operations at a large-scale. In general, due
to the complexity of the problem involved, large-scale system
optimization requires a non-trivial computational effort. In fact, the
time required to solve a normal energy dispatch problem increase
exponentially with the number of variable added to the problem \cite{quantum_comp_smartGrid}.
An example is provided by Cafaro \& Grossman in \cite{cafaro2014strategic} where a
mixed-integer non-linear programming model was created to optimize a
shale gas supply chain network. Here, the problem involves 50K variables
and 50K constrains for a computational time varying from 15 to 24 hours
depending on the calculation conditions. If the optimization is made
with a holistic approach, for instance, including transportation, energy
production, and demand, renewables generation, IoTs' fluxes of
information, the size of the problem, with respect to the computational
power of state-of-art supercomputers, simply would not be comparable.
Here, quantum computers come into play. `In 1982 Feynman
\cite{feynman1999simulating}~observed that quantum-mechanical systems have an
information-processing capability much greater than that of
corresponding classical systems, and could thus potentially be used to
implement a new type of powerful computer' \cite{jones1998implementation}. Here, differently
from a classical~computer that~encodes data into fundamental units
called `bits' whose state can be 1 or 0, a~quantum computer~encodes data
into `qubits' whose state can be 1, 0 or a combination. This is
practically translated in a dramatic improvement of the computational
power \cite{ladd2010quantum}, thus, the possibility to solve extra-large computational
optimization problems in a timely fashion, becomes realistic. In fact,
concerning the smart grid context, some proofs-of-concept have been
already provided to solve simplified problems, ranging from traffic flow
optimization to route optimization for multimodal transport systems
\cite{d_wave}. However, quantum computing is truly a game-changing technology
since, as previously stated, it will also likely push the boundaries of
cyber security and cryptography. Finally, in a smart grid perspective,
quantum computing will enable new paradigms in the energy market by
effectively preserving users' privacy and their economic transactions.

\subsection{Demand response \&
aggregators}\label{demand-response-aggregators}

Historically, energy demand and production matching i.e. load scheduling
problem, is one of the main challenges that systems operators have been
dealing with. Renewables' penetration has made this issue even more
challenging, due to the intermittent nature of these technologies.
Today's ICT allows employing demand response energy management systems,
whose scope is to control the energy demand to match the available
energy resources without adding new generation capacity \cite{haider2016review}. Today,
demand response can be applied also to the residential sector. Here, the
presence of highly connected home appliances i.e. IoTs will enable a
performing communication that is fundamental for controlling and
optimizing the energy system in a holistic and proactive fashion. To
this aim, aggregators technology is a key element in the communication
between operators and end-users (figure 5). These enable two-way communication to
achieve peak-shaving by modifying end-users consumption patterns
\cite{gkatzikis2013role}, thus optimizing energy consumption from one hand, and, energy
production on the other. In \cite{good2017review}, the authors provided a review of
the energy management systems aggregators highlighting the principal
gaps at technological, privacy and regulatory level. Precisely, from
this review study, it emerges the need for a highly efficient ICT
infrastructure, which must be associated with IoT, in order to properly
interact with end-users, for managing and balancing the energy
production and demand. Since these systems should have access to a broad
set of information to let the system work at its best, the authors
underlined the need to pay special attention to privacy issues. In fact,
high resolution metering data e.g. home appliances energy consumption,
represent a potential risk for privacy violation due to indirect and
implicit information carried, which could be easily triangulated by
third parties. Finally, the authors identified the need for an adequate
regulatory framework for demand response systems, playing a fundamental
role in the energy market and energy balancing. If these aspects are
considered and properly addressed, possibly through strategic
cooperation between industries, policymakers, research institutions,
demand-side management systems, and aggregators are likely to be the
strategic dowel in the smart grid puzzle to achieve a flexible and
efficient interconnected energy infrastructure.

\begin{figure}
\begin{center}
    \includegraphics[width=0.7\textwidth]{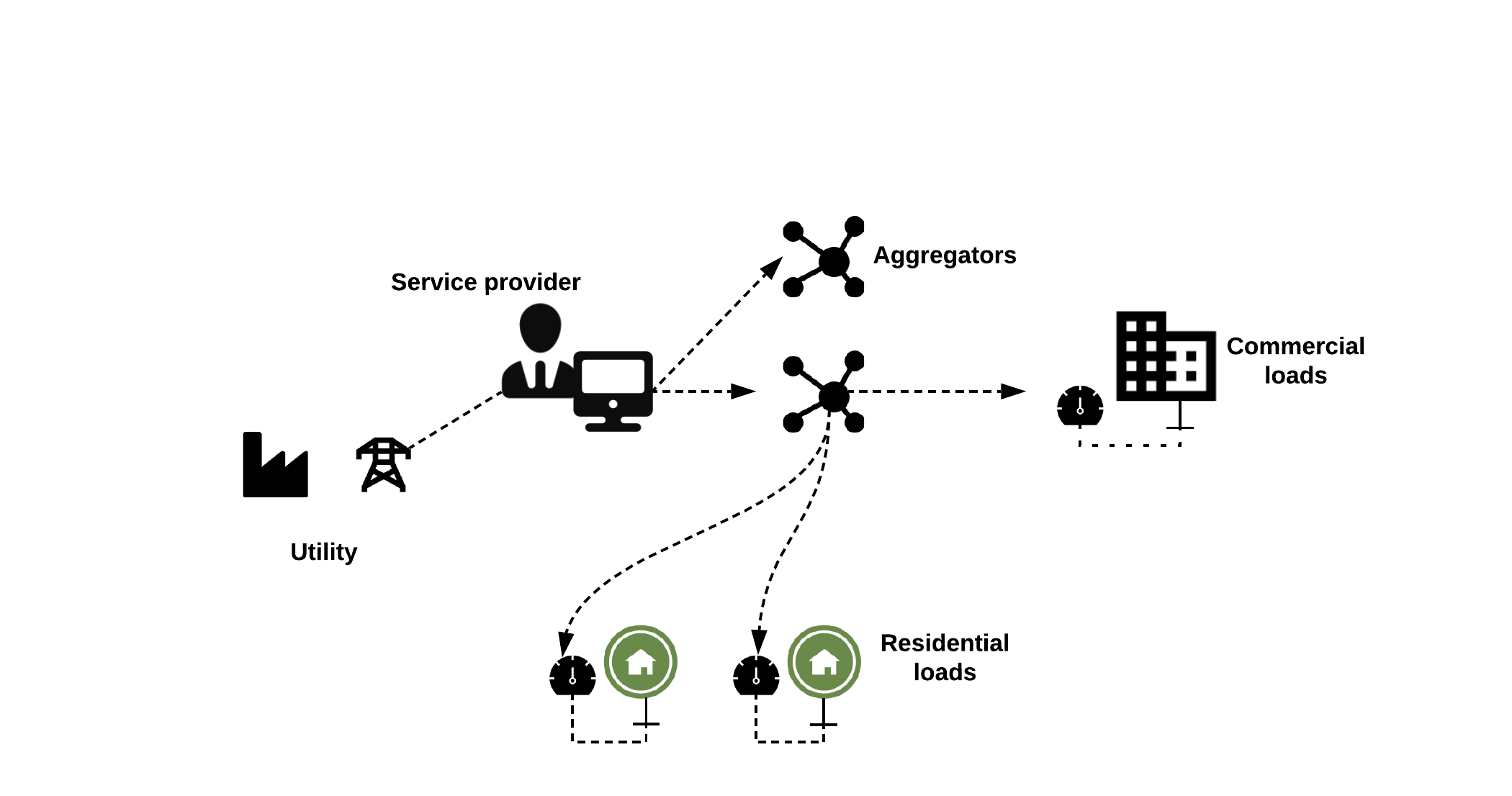}
\end{center}
\caption{A sample of demand-response application scheme \cite{elma2017overview}.}
\end{figure}

\subsection{Cloud computing}\label{cloud-computing}

Cloud computing provides large-scale integrated processing capabilities
which are more economically sustainable \cite{cloud}. In \cite{fang2016contributions}, Fang et
al. discussed the role of cloud computing within the smart grid
framework, identifying this technology as a potential solution to
mitigate disasters, increasing resiliency to large-scale failure. If
this last aspect is true from one side, from the other side, data
centers have to deal with different categories of risks ranging from
regulatory, technological, political to climate/natural. Precisely,
electricity blackouts, tornado/hurricanes, fires, flooding, earthquakes
or, unexpected events \cite{datacenters}. For this purpose, data center operators
employ different strategies for business continuity and disaster
recovery. These are mainly based on redundancy i.e. data are stored in
different locations (figure 6) and, from a business organizational level, an
integrated approach to manage problems is employed to enable business
resiliency from attacks or natural disasters. There is another problem
to deal with, in fact, data center providers need to take precautions to
prevent damages resulting from mid-size solar flares. This involves the
use of transient voltage surge suppression, uninterruptible power
supplies, on-site emergency standby generators \cite{solarflares2}. Finally, to
mitigate system outages by increasing network resiliency, the so-called
`fog computing' can be a strategic paradigm. Fog computing, compared to
cloud computing, favorites end-users proximity and it has a wider
geographical distribution \cite{bonomi2012fog}. In fact, in contraposition with the
cloud computing, where data are stored in a remote physical center, fog
computing foresees the use of more proxime devices, usually called edge
devices, to enable data storage and digital services.

\begin{figure}
\begin{center}
    \includegraphics[width=0.5\textwidth]{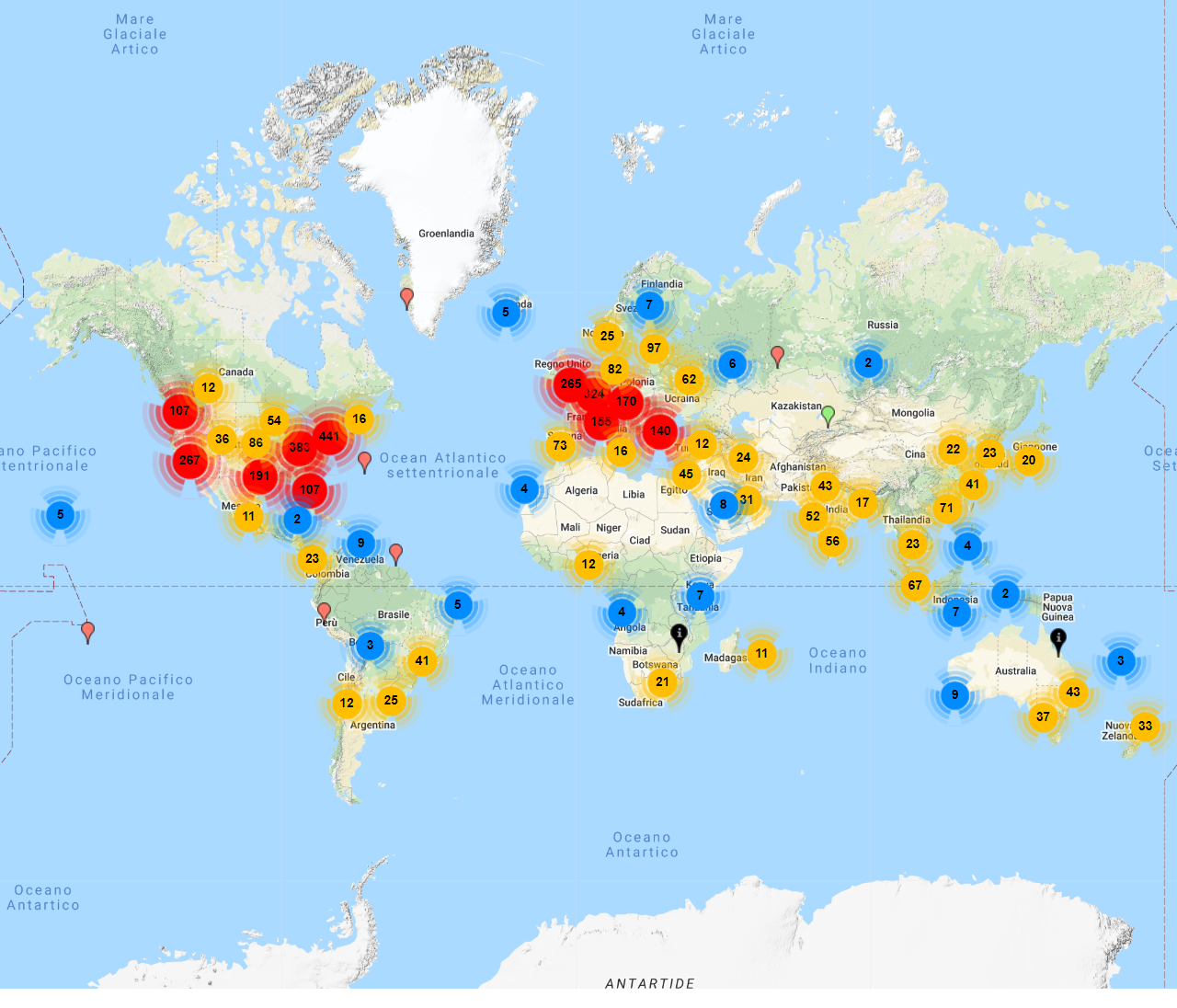}
\end{center}
\caption{Datacenter map \cite{datacenters_map}.}
\end{figure}

\subsection{Communication protocols }\label{communication-protocols}

Communication protocols refer to the set of rules that enable different
entities of a communication system to share information through
variations of physical quantities. The protocol comprises the rules,
syntax, semantics and, synchronization of the communication \cite{myers2001asynchronous}. In
2012, Usman and Shami \cite{usman2013evolution} discussed some of the major communication
protocols such as \emph{ZigBee} and \emph{WiMAX}, with a specific focus
on their application in smart grids and, as stated by the authors,
``smart devices have started to reach the consumer market but the
interoperability and complete solution for smart grid environment is
still far away''. In table 1, an overview of the main communication protocols is presented along with technical characteristics.

\begin{table}
    \centering
    \resizebox{14cm}{!}{
        \begin{tabular}{c|c|c|c|c|c|c|c}
        Feature & ZigBee/IEEE & Bluetooth/IEEE & Wi-Fi/IEEE & RFID & 12C & SPI & HomePlug 1.0 (PLC)  \\
         Base data Rate & 250 kbps & 1 Mbps & 11,000+ kbps & - & 100 kbps-3.4 Mbps & 20 Mbps & 14 Mbps \\ 
         Frequency & 2.45 GHz & 2.45 GHz & 2.45 GHz & 120 kHz - 10 GHz & lim. to 100 kHz, 400 kHz or 3.4 MHz & Free (n MHz to 10n MHz) : where n is an integer from 1 to 9 & 5000 kHz - 1 MHz \\
         Range & 10-100 m & 10 m & 1-100 m & 10cm - 200m & few meters & 100 m & 1-3 km \\
         Latency & 30 msec & 18-21 msec & 0.3 usec & 25-300 usec & Depends on the master clock & Depends on the masted clock & x \\
         Nodes/Masters & 65540 & 7 & 32 & - & 1024 & 2-3 & x \\
         Battery life & years & days & hours & battery-less & low power requirement & low power requirement & low \\
         Complexity & simple & complex & very complex & simple & simple hardware & simple hardware & simple \\
         Security & 128 bit & 128 bit & WPA/WPA2 & AES 128-bit & X & X & X 
    \end{tabular}}
    \caption{Communication protocols \cite{hafeez2014smart}}
    \label{tab:my_label}
\end{table}

\subsection{Prosumers}\label{prosumers}

Prosumer refers to a player which is involved in the production and
utilization of a generic good and it can be translated in ``production
by consumers''. Concerning the energy framework, Parag and Sovacool
\cite{parag2016electricity} watch at the prosumers paradigms identifying three different
categories: a peer-to-peer model where agents are interconnected,
prosumers-to-grid and prosumers community groups. The authors outlined a
possible successful scenario for prosumers' integration in the energy
market which could improve residential and commercial energy efficiency,
democratize demand-response and prepare society for distributed clean
energy technologies. However, the great market design is needed at
different levels otherwise, it could easily undermine grid reliability,
erode sensitive protections on privacy and inflate expectations to the
degree that the prosumer revolution satisfies nobody \cite{parag2016electricity}.

\subsection{Complementary applications, sectorial integration \&
synergies, technology
frontiers}\label{complementary-applications-sectorial-integration-synergies-technology-frontiers}

In addition to the abovementioned paradigms and technologies, there is
plenty of further innovative energy applications, management strategies
and emerging solutions that, in some form and in some way, will
characterize the portrait of the smart grid of the future. For example,
without the aim of exhaustiveness, vehicles-to-grid (V2G) and battery swapping
applications (P2G) \cite{kempton2005vehicle} are complementary paradigms that will take part in
the smart grid shaping process for some contexts. Similarly,
power-to-gas applications \cite{gotz2016renewable} is another current issue that
scientists are dealing with. Also, energy storage, in a broader sense of
the term, and sectorial integration i.e. industrial symbiosis, waste
heat recovery, is to increase the flexibility and sustainability of
energy systems operations, affecting decisively, the evolution of our
technological landscape for the energy context. Finally, some minor
applications such as energy recovery from natural gas distribution (ER)
\cite{cascio2018optimal} and emerging control strategies such as gas-bagging \cite{cascio2018flexible}
applications, in a long term perspective are likely to contribute to
shaping the smart grid scenario as well. Or, for the sake of ontological
coherency, the smart grid scenario, intended as a whole, is likely to
shape the contribution of these applications. Besides, the smart grid of
the future will be likely characterized by frontier technologies that
are currently being studied or developed. For instance, researchers are
developing a technology to convert a wall into a trackpad and motion
sensor and this could be achieved thanks to a conductive paint \cite{wall_paint}.
Once this technology will reach a certain level of matureness, smart
walls will be presumably able to track people's gestures or monitor
appliances. As regards this aspect, it comes intuitively to understand
the potential level of insights that could be achieved by monitoring
people's body language, gestures and so on. Further aspects affecting
the smart grid of the future could reside in complementary sectors and
their technological advances. For instance, the space exploration and
colonization sector have synergies with the smart grid sector. In fact,
``NASA and smart grid both need autonomous controls'' \cite{nasa}. A
further practical example of intersectoral synergy can be represented by
the \emph{SpaceX Starlink} project. This consists of a constellation of
thousands of mass-produced small satellites working in combination with
ground transceivers, to provide broad internet access, thus improving
smart grid applications performance, making it easy to implement smart
grid technologies also in remote areas.

\subsection{Artificial intelligence }\label{artificial-intelligence}

Artificial intelligence and machine learning are increasingly seen as
key technologies for building more decentralized and resilient energy
grids. These techniques are  powerful tools for design, simulation, control, estimation, fault
diagnostics, and fault-tolerant control in the smart grid \cite{bose2017artificial}.
However, their development needs to be properly addressed. Some
researchers emphasize the need to consider the ethical and social
implications of these developments \cite{robu2019consider}, thus, artificial
intelligence framework should pass through a regulatory process to
enable sustainable development, otherwise, it could result in gaps in
transparency, safety, and ethical standards \cite{vinuesa2020role}. But, from a
technological perspective, as reported in \cite{strukov2019building}, the artificial
intelligence has made such huge steps forward that we have arrived at a
scientific frontier where -- citing the authors -- `artificial
intelligence needs new hardware, not just new algorithm'. This brought
some scientists to focus on the possibility of building brain-inspired
computing \cite{strukov2019building}. Precisely, the idea of the so-called neuromorphic
computing is to design computer chips inspired to the brain, thus
merging memory and processing units, achieving impressive computational
power and speed with very little power consumption. This will enable
complex deep learning networks functioning that would help to solve --
in a prompt fashion -- complex problems related to the smart grid
control.

\subsection{Big Data}\label{forecasting}

Smart sensors networks are a great opportunity for smart grid applications due to the high level of magnitude of data gathering. However, it also brings new challenges and costs for storing and processing consistent flows of information with a high frequency \cite{jaradat2015internet}, which are commonly identified with the term `big data'. Precisely, as report in \cite{bibri2017ict}, `big data' universe involves the use of tools (e.g. classification, clustering, and regression algorithms), techniques (e.g. data mining, machine learning, and statistical analysis), and technologies (e.g. Hadoop, Hbase, and MongoDB) that are used to extract useful knowledge from large fluxes of data. In \cite{tu2017big}, the authors reviewed the big data issues for the smart grid framework, highlighting challenges and opportunities. At 2017, the authors believes that the big data domain, even thought it is rapidly leaving, still it is in a early stage and, in a future perspective, different technological points should be faced. And these are:
\begin{itemize}
\item multi-source data integration and storage,
\item  real-time data processing, data compression,
\item  big data visualization, 
\item and data privacy and security.      
\end{itemize}     

\section{Regulation, security and social factors
  }\label{regulation-security-and-social-factors}

 \subsection{Regulation, privacy \& cyber security
    }\label{regulation-privacy-cyberg-security}

According to Iqtiyaniilham et al. \cite{iqtiyaniilham2017european}, the European Union maintains
world leadership in smart grid technology. The authors identify the
integration of various disciplines, overcoming regulatory barriers,
technology maturity, and consumer engagement as the key challenges for
those experts involved in disciplines gravitating around the smart grid.
This is valid especially for international operators such as the
European Network of Transmission System Operators for Electricity
(ENTSO-E) and the Coordination of Electricity System Operators (CORESO)
which are the agencies that presently coordinate system interconnection
and operation \cite{iqtiyaniilham2017european}. The orchestration of the interdisciplinary
problem has been recently faced by the European Commission by
instituting a Smart Grid Task Force which comprises different subgroups.
For instance, one group is focusing on cyber security and it is to
prepare the ground for sector-specific rules for cyber security aspects
of cross-border electricity flows, on common minimum requirements,
planning, monitoring, reporting, and crisis management for the
electricity subsector \cite{taskforce1}. Other groups, instead, focuses on the
deployment of demand-side flexibility and the specific case of explicit
demand response in Europe \cite{taskforce3}.\\
The European Commission, in 2019, adopted specific guidance
(\emph{recommendation C(2019)240 final and SWD(2019) 1240 final}) to
implement cyber security rules with the final aim to improve awareness
and organization in the energy sector \cite{cybersecurity}. However, historically,
technology and regulations travel at different speeds and, concerning
the digital e-cosmo, technologies here are often put on the market
without proper comprehension of the privacy and cyber security risks. As
highlighted in \cite{mylrea2017smart}, the technologies involved in the smart-grid
landscape, will have to understand how the information is collected,
stored, sold, used, and what jurisdictions does the information
traverse. This challenge it's far from being trivial since, as clearly
explained by Edward J. Snowden in \cite{anderson2019edward}, it is necessary to identify
security measures for privacy protection which have to involve both data
and metadata linked to users energy usage and other domains. Precisely,
regulations will have to be designed and implemented with a resolutive
approach, facing in a clear, explicit and non-ambiguous way the
jurisprudentia relative to privacy violations due to business
intelligence triangulations based on metadata-type-information. Besides,
the ``flip side'' of the digital innovation does not only involve
privacy. In fact, in the smart grid framework, Khatoun and Zeadally
\cite{khatoun2017cybersecurity} identify privacy and public safety as a priority for political
debate and scientific research, highlighting the imperative need to
contrast cyber crime in smart cities for every class of cyber-attack:
cyber warfare, terrorism, industrial espionage, activism, economic
reasons to jokes \cite{otuoze2018smart}.

\subsection{Behavioural aspects and people's perception
}\label{behavioural-aspects-and-peoples-perception}

As reported in \cite{bigerna2015overview}, multidisciplinary cooperation is needed to
develop scientific research on smart grids since the creation of new
infrastructures is generally linked to acceptance problems which are
important for the adoption of new technologies. In this regard, the
inclusion of conscious consumers in the process is a fundamental issue
to be addressed in the smart grid realization where the problem involved
is extremely complex due to its multidisciplinary nature. In fact, there
is plenty of social and psychological issues gravitating around the
smart grid concept. These range from human factors \& energy consumption
to effect of technology on society, also in a broader sense of the term,
thus involving issues at the anthropological level as well. Precisely,
energy consumption in buildings, especially residential, largely depends
on human behaviour and the context in which energy-relevant decisions
are being made, thus social sciences and IT could benefit from each
other \cite{tiefenbeck2017bring}. In this regard, De Dominicis et al. \cite{de2019making} analyzed
the impact of real-time feedback on residential electricity consumption,
highlighting how feedbacks based on social comparison resulted in a
sensible long-term reduction. Tiefenbeck et al. \cite{tiefenbeck2019real} proposed a case
study based on 6 hotels monitoring -- 265 rooms, 19.596 observations --
demonstrating how sensible is the behavioural factors in energy
consumption. Precisely, this study proved that a real-time feedback
intervention would result in a considerable 11.4\% energy reduction.
Furthermore, gamification i.e. the use of game mechanics to drive
engagement is a reinforcing factor that has been shown to encourage
targeted behaviours with instant positive feedback \cite{bradley2011review}.\\
At a societal level instead, Boudedt in \cite{boudet2019public} reviewed the literature
on public perceptions relative to different energy technologies from a
broader perspective. The author concludes highlighting how research on
public perceptions relative to new energy technologies will have to
continue to shed valuable light on the complex interface between energy
technologies and the broader society they serve. In fact, as stated by
Norman \cite{norman2018autonomous}, cities exist within a wider system and it may take more
than technological advances, innovation, and city autonomy to develop a
sustainable urban future.

\subsection{Technical risks for the smart
grid}\label{technical-risks-for-the-smart-grid}

In general, the smart grid is to all effect a set of multi-domain
interdependent networks of systems and human players, and this makes the
smart grid -- whatever is its configuration -- subjected to those common
risks that are typically manifested in all complex networks: cascade
failures. Failures have been studied especially in the electrical
transmission domain and, overload failures usually propagate through
collective interactions among system components and the propagation
dynamics of the cascading failures are essentially unknown \cite{daqing2014spatial}.
Buldyrev et al. \cite{buldyrev2010catastrophic} developed a framework for understanding the
robustness of interacting networks subject to such cascading failures.
From the findings of this study, it emerges the need to consider
interdependent network properties in designing robust networks. In fact,
according to the authors, a broader degree distribution increases the
vulnerability of interdependent networks to random failure, which is
opposite to how a single network behaves. The same needs at the systems'
design level, has been highlighted in \cite{vespignani2010fragility} by Vespignani. For the
electrical transportation infrastructure, some efforts have been made to
prevent power outages. For example, there is a technology called
synchrophasors which are equipped with GPS, this technology gives you
microsecond accuracy of time across the whole power system \cite{haq2017smart}.
Transient dynamic behaviours for dynamically induced cascade failures
have been studied by Schäfe et al. in \cite{schafer2018dynamically}, highlighting the need
for further investigation to outline failures propagation dynamics and
mitigation strategies. In \cite{robu2019consider}, Robu et al. wisely highlighted the
fundamental ethical and social challenges for the digital revolution
which is irreversibly shaping the smart grid scenario. Conclusions
suggest the need for a careful control for the design and realization of
the smart grids, whose increasing architectural complexity and AI need
to be properly ensured, to prevent, for instance, drastic blackouts.
Thus, except for the presumed reduction of the pollution level \cite{ball2004blackout},
blackouts might be responsible for generating several dramatic issues
and its prevention is an unquestionable priority. Among all the possible
causes, blackouts may be originated also by solar flares. Solar flares
produce high energy particles (primarily high-energy protons) and
radiation (primarily x-rays). To be more precise, this last disturb the
ionosphere -- from 9 to 200 km -- affecting radio communications.
Besides, along with energetic ultraviolet radiation, they heat the
Earth's outer atmosphere, causing it to expand. This increases the drag
on Earth-orbiting satellites, reducing their lifetime in orbit \cite{solarflares}
or damaging astronauts in orbit \cite{solar_storm_effect}, or cause severe damage to
electrical systems and communications \cite{morina2019probability}. Thus, both intense radio
emissions from flares and changes in the atmosphere can degrade
satellite communications, especially for Global Positioning System (GPS)
measurements \cite{solarflares}. Along with solar flares, scientists now
understood that the major geomagnetic storms are induced by coronal mass
ejections and this are frequently associated with flares. Coronal mass
ejections likely have a 11-years cycle. There is a serious problem
associated with geomagnetic storms that is the damage of Earth-orbiting
satellites, especially those in high, geosynchronous orbits. As reported
in \cite{solarflares}, in 1989 high currents in magnetosphere induced high currents
in power lines, blowing out electric transformers and power stations
and, this risk involves mainly high altitude zones, where induced
currents are greatest, and those areas having long power lines and/or
where ground is poorly conducting \cite{solarflares}. Due to this risk, scientists
are exploring the possibility to predict Carrington events. In this
perspective, some authors estimated that the probability of occurrence
of an 1859-Carrington-like event is estimated to be between 0.46\% and
1.88\% - much lower than what is identified in the literature. While, on
the other hand, some scientists are even proposing to protect Earth
from solar flares \cite{lingam2017impact} by employing a magnetic field to deflect
charged particles (figure 7). The authors approached the feasibility of the measure
from a physical perspective, in terms of its basic physical parameters,
highlighting no specific limitations. This research carry-out a heuristic
analysis of the potential economic impact of such an extreme space
weather event, comparing the cost of the mission for lifting a
10\textsuperscript{5} tons object into space, would be around \$100
billion -- assuming a \$1000 per kg -- which is comparable to the total
cost of the International Space Station, but still 3-4 orders of
magnitude lower than the economic loss generated by a catastrophic event
\cite{lingam2017impact}, that would directly compromise the smart grid.

\begin{figure}[ht]
\begin{center}
    \includegraphics[width=0.5\textwidth]{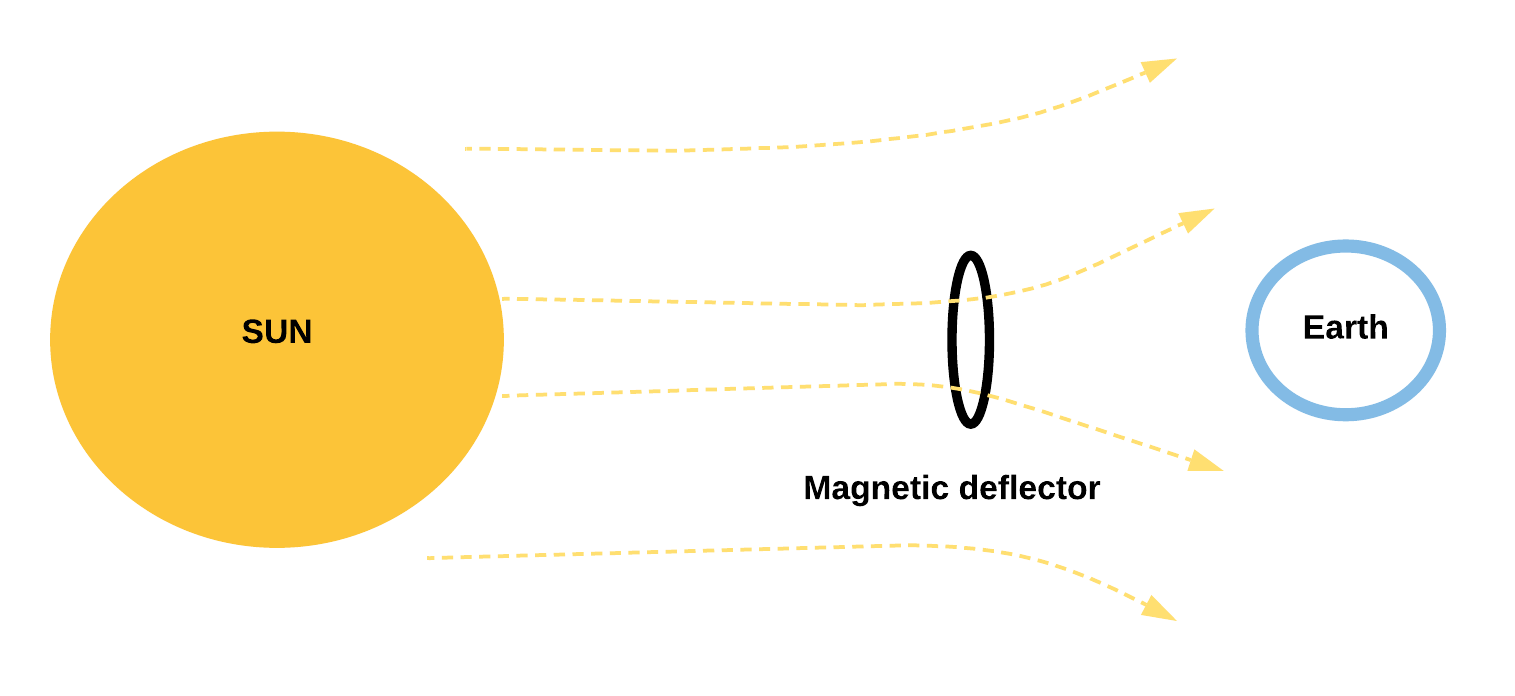}
\end{center}
\caption{Earth magnetic shield \cite{lingam2017impact}.}
\end{figure}

\section{How smart is the grid}\label{discussion}

In this section, based on the smart grid anthology presented, an idealization of the smart grid concept is carried out in order to support, first, the definition of a taxonomic framework for smart grid assessment, secondly, a discussion relative to the ontology of the smart grid and its transition (figure 8), thus to focalise the premises behind the realization of the smart grid concept.

\begin{figure}
\begin{center}
    \includegraphics[width=0.5\textwidth]{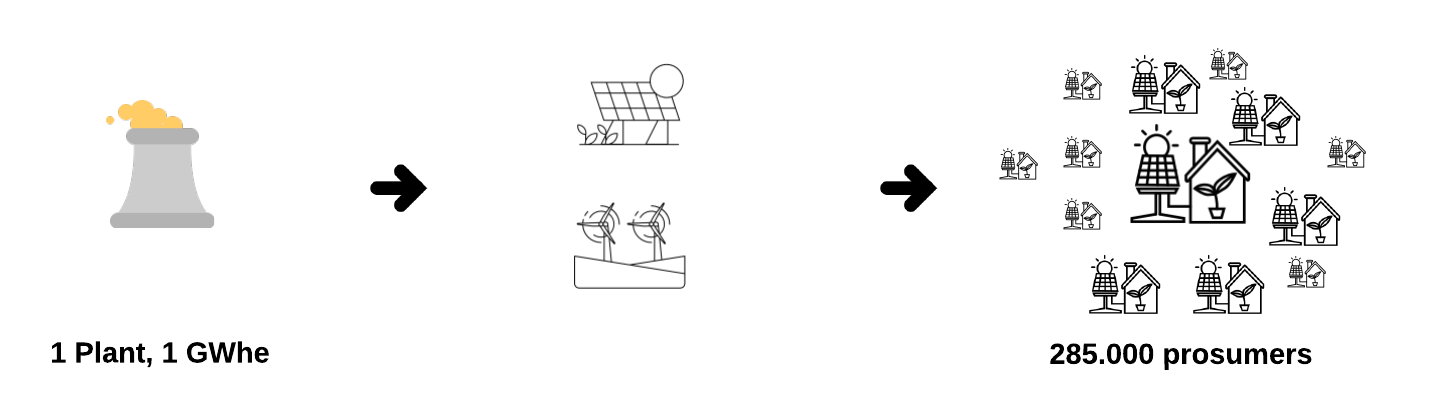}
\end{center}
\caption{Paradigm shift: from central to distributed generation}
\end{figure}

\subsection{Subjects and objects }\label{subjects-and-objects}

To focalize the level of complexity that will likely characterize the
current and the emerging smart grid, it would be helpful to make few steps backward,
trying to identify from a distant perspective, which are the main
elements, interconnections, layers, and actors that will come into play,
defining the presumable final portrait of the whole set. To this aim, in
figure 9, a simplified scheme of an idealized smart grid is presented. Concerning figure 9, the smart grid sub-borders can be
ideally divided into resources, supply, storage, and harvesting.  Concerning the energy vectors, electricity, natural gas, thermal energy, hydrogen networks, other
renewable carriers and eventually non-renewable carriers e.g. oil.
Finally, figure 9 includes the ICT overall infrastructure and enabled paradigms (cloud and
fog computing and demand-side management etc). These are linked to a complementary layer which ideally involves issues concerning the economic, financial
and regulation dynamics, along with social one.

\begin{figure}[ht]
\begin{center}
    \includegraphics[width=0.9\textwidth]{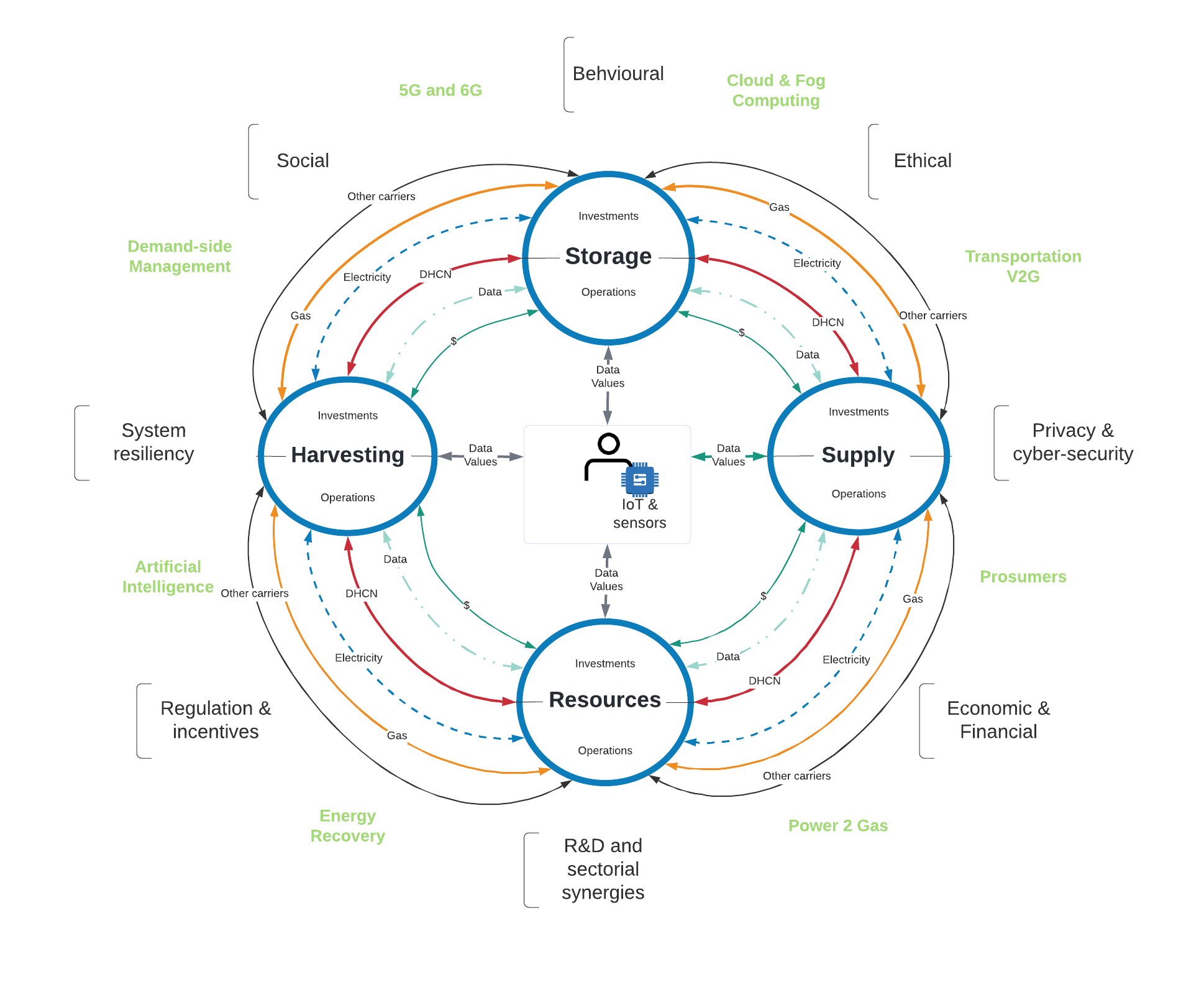}
\end{center}
\caption{An idealization of smart grid universe.}
\end{figure}

\subsection{Synergies vs complexity}\label{synergies-vs-complexity}

The main issue emerging from figure 9 depicting the idealized smart grid universe,
concerns the dichotomy `synergy vs complexity' which derives from the
multisectoral integration that digitalization and technological progress
has made enabled. In fact, on the one hand, the sectorial integration
makes it possible to give rise to synergies that would most likely be
translated in energy, economic and environmental benefits, which have
been thoroughly discussed above. On the other hand, instead,
digitalization is constantly increasing the level complexity of the grid
enabling a high level of sectorial interdependency. Thus, without the
aim of exhaustiveness, we can in general state that, depending on the
system architecture, energy service continuity and performance in the
next future might intimately be linked to information deriving from IoTs
environments, which is in turn subjected to ICT infrastructure
functioning and/or GPS services, which in turn are dependent to
satellites integrity. At this stage, the main concern outlined from this
study involves our lack of comprehension and ability to predict the
level of resiliency of the emerging smart grids that, moreover, varies
from case to case. In this sense, the possibility that system integration and sophisticated
technological solutions might not be the best compromise for large-scale
systems, should be taken into account. In fact, system complexity foreseen being subjected to a higher
number of risks or, in other words, citing Ford, ``what is not there,
does not break''. To be more precise, considering how the smart grid
tends to be designed so far, it should not be excluded the possibility
that an extreme event, such as Carrington event, or a set of
non-predictable circumstances -- as the dramatic Fukushima disaster
teach us -- could generate a sequence of large-scale failures which
could likely evolve on the mid-term, compromising communication, energy
services continuity, and reflecting on the society by perturbating, in a
non-desirable way, economic equilibriums and social dynamics.

\subsection{Taxonomy \& categorization}\label{tax-cat}

Starting from the smart grid anthology carried out in the previous sections, a preliminary taxonomic framework for smart grid categorization can be  proposed. This framework is conceptually based on the  analysis of the technological evolution developed through three different progressive generations of the technological maturity and the properties of a grid (figure 10). First-generation grids can be characterized by less sophisticated and less advanced technological content and is mainly  targeting centralised production systems based  on transmission  lines: e.g. electric transmission infrastructure equipped with ICT for power quality monitoring and control. Second generation grids is considering more decentralised production and consumption, they are characterized by a lowest common denominator which is represented by a distributed-based generation. Finally, third generation grids would embed the characteristics of the previous ones while they are considered to be the most sophisticated as they would foresee the use of "pseudo-empathic" features where, for example, users engagement is enabled through AI-based tailored feedback and effective dynamically regulated end-users communication.
Also, it is worth highlighting how the term `smart grid' assumes a broader meaning, becoming at a certain point a synonymous of `smart city' or `smart energy system', as the domain's borders merge  with  complementary domains.  In figure 10, a systematic decomposition is applied to outline a smart grid framework based on the above mentioned three generations (1st, 2nd, and 3rd), while each generation is identified by a generic taxonomy. These are following discussed in details. 
\subsubsection{First Generation}
\begin{itemize}
\item
  \emph{Integrated.} the simplest possible instance of an integrated energy system can be represented by a transmission network equipped with ICT technologies for monitoring and power quality control. However, in the reality, the integration concept might assume a broader meaning, thus involving other energy carriers, including water, gas and thermal networks, from the production side, embracing residential and tertiary sectors through IoTs technologies, for the end-users side. As discussed in the previous section, energy system integration is the key to achieve higher operational flexibility thus to theoretically unlock the systems' potential at a large scale, in terms of operational optimization and renewable resources penetration. When integration occurs at wider level, it can involve also transportation sector or industry e.g. vehicles to grid application or industrial symbiosis. However, system integration is not always synonymous of optimized operations, as it should be intended as a necessary but not sufficient condition to achieve a higher level of sustainability.
 \end{itemize}
 
 \begin{itemize}
\item
  \emph{Optimized.} Energy systems optimization involves the conjunct resolution of rigorous and complex mathematical problems across three different levels: system configuration (synthesis problem), design (component size) and operation (system control). These levels are not "watertight compartments" as, for instance, optimal design could be related to control and vice-versa. Similarly, even if the optimization is conducted through rigorous mathematical models, it remains a relative concepts whose result can varies depending on the dimension of the conceptual borders and study premises: building level, district level, regional or national level. Of course, in this perspective, to be characterized as "optimized" the energy systems architecture, design and control, should be desirably optimized in a long-term perspective, thus involving also aspects such as climate change resiliency and anthropological viability for a sustainable inheriting for future generations.         
 \end{itemize}
 
 \begin{itemize}
\item
  \emph{Secure.} As discussed in the previous sections, information and communication technologies are revolutionizing the energy landscape, enabling new positive paradigms and business models. However, ICT brings new challenges for privacy and cyber-security which are destined to call the society to develop novel frameworks and technologies that concretely allow us to create a "secure" energy system. From a privacy perspective, it would be necessary, for a secure energy system to employ high data and metadata protection standards, being at the same time resilient to cyber-attacks. But, "security" goes farther than the ICT domain, as it also involves the concept of system resilience and homeostatic features to environment changes and to drastic, unexpected, and extreme scenarios. For instance, and without the aim of completeness, these might include resiliency to pandemics, tornado, earthquakes, terrorism, etc, foreseeing also extreme space weather events, high altitude orbits monitoring and space crime security. As regard to these last points, energy system design policies should be harmonized with respect to space programs and, last but not least, telecommunication technologies, with specific reference to 5G and 6G should be thoroughly investigated from a health perspective before being deployed.     
 \end{itemize}
 \subsubsection{Second Generation}
 \begin{itemize}
\item
  \emph{Distributed.} Renewable penetration, energy technology accessibility, ICT revolution are some of those main factors that are shaping the energy conversion and management paradigm, letting it switch from centralized to distributed generation. Decentralization is likely going to positively affect different domains ranging from market dynamics, energy accessibility, to energy system resiliency. Especially on this last point, a heterogeneous and context-harmonized energy system is desirable, thus technological diversity should be encouraged also for the ICT domain, promoting fog-based computing for instance. The level of decentralization of electrical systems could also evolve in planetary grids, where renewable production is capitalized through power exchanges between different time zones (e.g. Europe, USA - China) \cite{monti2018global}. Finally, the design or retrofitting of a generic energy system should be done by favouring the decentralization but also considering the potential penetration of nuclear fusion. Nuclear fusion might be available for industrialized countries by the mid of this century \cite{Nuclear_fusion} and, how it will affect the renewable market and renewable technology penetration? Will we assist to a re-centralization of the energy production with nuclear fusion? With this in mind, energy system should be designed and developed on a long term perspective, trying to find a harmonized coexistence of technologies, to keep the objective benefits (commercial, resiliency etc) that the decentralization paradigm brings by its nature. 
  
 \end{itemize}
 
 \begin{itemize}
\item
  \emph{Democratized.} Decentralization is the antechamber of the ambitious challenges of what is called energy democratization. A distributed energy system with a widespread renewable generation "prepares the terrain" for enabling those paradigms which are likely to unlock energy accessibility at a global scale, with an undoubtedly positive impact on geopolitical equilibrium, smoothing/neutralizing resource monopolization. As discussed in the previous section, one of the most important technology for creating a "democratized" energy system is certainly the blockchain which will be capable to unlock, for instance, a reliable market framework for peer-to-peer energy exchange across energy communities and prosumer clusters.    
   
 \end{itemize}
 
  \begin{itemize}
\item
  \emph{Circular.} Circularity refers to the virtuous organization of a economic system, included an energy system, based on the reuse, sharing, repair, refurbishment, re-manufacturing and recycling to create a closed-loop system, minimising the use of resource inputs and the creation of waste, pollution and carbon emissions \cite{geissdoerfer2017circular}. Examples of virtuous energy systems could be represented by bio-fuels production and local exploitation, or netZero energy buildings as, for instance, the Australian \textit{Sustainable Buildings Research Centre} \cite{SBRC}.        
 \end{itemize}
 \subsubsection{Third Generation}
 \begin{itemize}
\item
  \emph{Interactive.} This taxonomy refers to the case where end-users are pro-actively engaged in the energy management process, thus, achieving a certain level of awareness through system interactions enabled by ICT and IoT technologies. In this case, a two-way communication and the possibility for the system to control the operational status of some IoTs from the demand side, becomes a fundamental prerequisite.         
 \end{itemize}
 
 \begin{itemize}
\item
  \emph{Semantic.} Semantic functionalities are possible when the energy systems are enough mature and sophisticated to enable real-time wide-area optimization (wide-area awareness), thus, involving energy exchanges between clusters of prosumers with fair game-theoretic based control. Also, the dynamic resiliency of the energy system is eventually assessed to ensure service continuity across the communities.     
 \end{itemize}
 
  \begin{itemize}
\item
  \emph{Pseudo-cognitive.} When digital technologies and artificial intelligence are strategically employed in energy systems management, advanced functionalities might be enabled, and these refers to self-healing, pseudo-emphatic end-users communication, and even e-stress monitoring and control. This refers to the possibility to generates customized and tailored automation based on users' habits, harmonizing the user engagement with respect to energy efficiency and psychological aspects as well, thus avoiding stress generated by excessive use and presence of digital technologies in the human environment.    
 \end{itemize}

Finally, it is important to highlight that the technological development process is heuristics and iterative by its nature rather than linear. And this means that, when applying this framework to the reality of today's technological smart grid scenario, third generation's features might be found in first or second generation grids for instance. On the other hand, it is true that this would make the categorization of a grid a challenging task and every labeling attempt might result to be not consistent after all. However, when this framework is employed in conjunction with a structured road map of propaedeutic actions designed for the specific context, classification uncertainty might be neutralized as, for example, achieving third-generation status might require the accomplishment and the integration of features of the previous generations (second and first). Finally, the level of smartness of a generic grid, even if objective criteria are employed, tends to result to be a more relative concept rather than objective, as there might be different possible compromises of first and third generation features for instance, that would result in an harmonized solution with respect to the local context and the territory priorities. In other words, for example, a first generation grid is characterized by a set of advantages (simplicity, resiliency etc) that might represent the smartest solution for that given context. Concluding, even it might result apparently paradoxical, it might be also true that first generation grids might not be synonymous of a less sustainable solution. Concluding, an fundamental aspect that should be taken into account when assessing the sustainability of a grid, is the amount of auxiliary energy (i.e. the eventual CO2 production) necessary for the conduction and operation of the grid itself (including ICT and IoTs related energy consumption) i.e. directly or indirectly associable to the presence of the energy system itself. In this perspective, it should be noted that it IoTs technologies is being widespreadly used  (smart city, agriculture, transportation, etc.) These devices will produce an important amount of e-waste while consuming an important amount of energy as well \cite{abedin2015system}, thus eventually generating carbon emissions.

\begin{figure}
\begin{center}
    \includegraphics[width=0.9\textwidth]{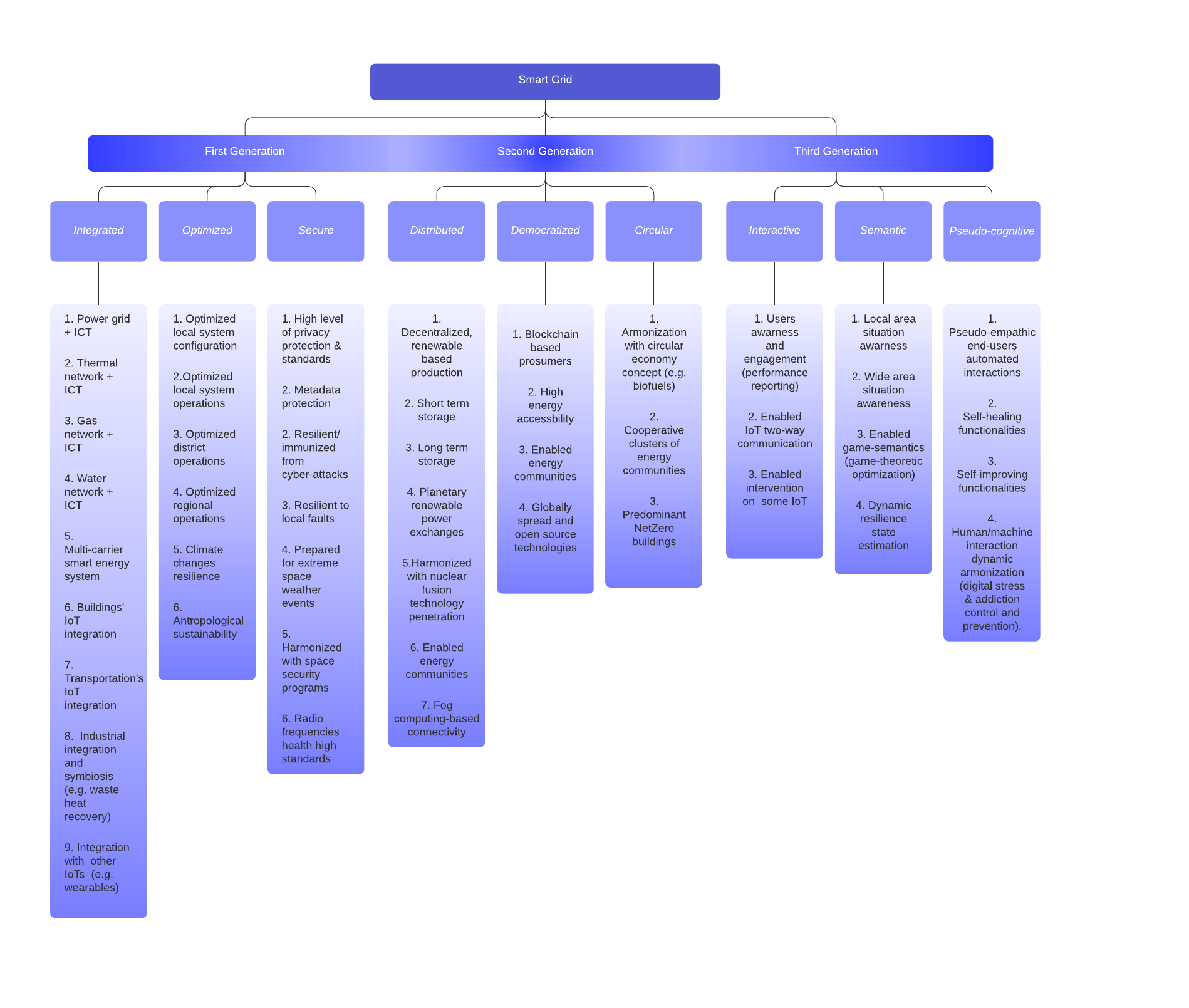}
\end{center}
\caption{Smart grid taxonomic framework and categorization.}
\end{figure}

\subsection{Next step?}\label{next-step}

Two levels of actions are identified as `next step' to increase the
chances of achieving a sustainable grid in the most general sense of the
term. These are mid-terms actions and long-term actions. Concerning the
mid-term, there are several issues that researchers, industries and
regulators should focus on, and these are:

\begin{itemize}
\item
  \emph{Ontological.} Study and debate of the ontological premises
  relative to the realization of the smart grid concept and its
  contours. Precisely, apart from the potential RETs share increase that
  digital technologies could enable, from a social, environmental and
  human-life-quality perspective, where the progress resides when
  creating a highly digitalized and interdependent grid? From this
  point, there are several sub-issues -- following presented -- that
  need to be properly and thoroughly discussed, avoiding leaving them to
  be pulled and shaped by market and commercial speculation.
\item
  \emph{Behavioural.} Cyber technologies offer new risks for new types
  of addiction \cite{takahashi2018behavioral}. Thus, it is important to understand social and
  psychological eventual risks deriving from a highly digitalized
  environment and surrounding (IoT) which, as previously described, are
  likely to be functional to the presence of a fully automated grid.
  This refers especially to people's homes.
\item
  \emph{Ethical.} `Permanent records' \cite{anderson2019edward} and sensible data storage
  risk to subtly erode our freedom -- in the most general sense of the
  term -- at its roots. In particular, for instance, the fact is that we
  have no idea about the final effect on individuals' physiology
  generated by the conscious awareness of the presence of `permanent
  records' or all those information gathers by digital technologies,
  including social media. Thus, to the best of our knowledge, we cannot
  exclude the possibility that these may be a serious ontological bug in
  our society which is likely to be ignored by the most due to its
  subtle and non-measurable effects. In this sense, from a general
  perspective, the evolution of the smart grid will drastically increase
  the e-traffic, thus its development must be designed (or retrofitted)
  by paying particular attention to users' privacy to prevent invisible
  societal disasters whose eventual existence, is unexplored terrain.
\item
  \emph{Technical.} Systematic studies are desiderable to understand
  complex dynamics and emergent behaviours of interdependent systems
  (ICT, energy systems, IoT, and complementary associated sectors e.g.
  transportation and autonomous vehicles). These can be carried out
  through large-scale numerical simulations, through dedicated research
  programs.
\item
  \emph{Regulations.} Normative should be designed to rigorously assess
  final energy benefits (if any) generated by the employment of
  different digital measures and paradigms at different levels: urban
  level e.g. aggregators, building level e.g. electronic `smart' devices
  for HVAC control, etc. Thus, the design of new incentives to unlock
  and regulate new business models such as peer-to-peer energy exchange
  is very recommendable to sustain energy communities, increase system
  resiliency.
\item
  \emph{Research \& training.} Increase the number of strategic research programs to
  enrich knowledge, generate highly trained professionals to enable
  dedicated consultancy to design tailored solutions (through digital
  twins for instance) to approach the unavoidable architectural
  heterogeneity of the emerging smart grids.
\end{itemize}

On the long-term perspective instead, there are further aspects that
directly or indirectly are linked to the smart grid universe. These
range from space-security i.e. space crime prevention e.g. satellites
kidnapping, to missions and projects to explore the Sun to increase our
comprehension of the star cycles and extreme solar events forecasting,
to prevent eventual drastic blackouts, transversal network cascade
failures, and vertical inter-sectorial domino effects. Also, it might be necessary to re-think and harmonize the eventual penetration roadmaps of nuclear fusion technologies with respect to the distributed generation concept and the advantages derived from this, thus neutralizing the risk of a `re-centralization' of the energy production that nuclear fusion technology would likely be able to induce due to its game changing potential.  

\section{Conclusions
}\label{conclusions-how-smart-is-the-grid}

In this article, a systematic transversal review of the emerging
paradigms for the smart grid was presented. From a technological
perspective, different paradigms have been discussed and contextualized
with respect to the smart grid framework, identifying synergies and
limitations. The paradigms considered are IoTs, smart meters,
blockchain, 5G/6G, quantum computing, demand response, cloud \& fog
computing, communication protocols, prosumers, artificial intelligence,
and further complementary applications. From this smart grid anthology, a taxonomic model for smart grid categorization was outlined. This involves three different generations (1st, 2nd, 3rd), while each
generation is identified by a generic taxonomy i.e. integrated, optimized, secure, (1st) distributed, democratized, circular, (2nd) interactive, semantic and pseudo-cognitive (3rd). Furthermore, from the scenario portrayed, a
set of issues involving the regulation, security, and social frameworks
have been further derived in a theoretical fashion. Thus, from an
engineering perspective, the presumed most critical issue outlined
involves the dichotomy `synergy vs complexity', which derives from the
multisectoral integration that digitalization and technological progress
has made enabled. Precisely, this refers to our lack of comprehension of
faults propagation mechanisms, thus the level of resiliency of the
emerging smart grids. For instance, the highest risk deriving from an
interdependent highly digitalized grid, might be represented by the
possibility of a Carrington event or a set of non-predictable,
non-desirable circumstances. 
Finally, the engineering challenges and the mitigation/neutralization of these risks can be mostly addressed by employing a nature-inspired development approach: the human body for instance, is the best definition of a extremely complex but sustainable system, i.e. it is a truly smart system. Thus, the energy systems of our future societies, should - and most likely will - be inspired to the human body intended as a homeostatic system involving energy supplying, managing, storing, harvesting and control, and these features will likely characterize the taxonomy of the fourth generation smart grids.  
Besides, the technological framework which
is developed is also intimately linked to socio-anthropological
aspects that are identified, without the aim of completeness, in this
review. As regards these, the most critical aspect outlined through this
study resides in the risk of an "ontological misunderstanding"
relative to the realization of the smart grid and its contours, which could likely occur during the development path. In fact, at a global level,
from a human-life-quality perspective, it's known that the main
advantage deriving from grid digitalization is linked to the possibility
to increase the flexibility of the system, thus increasing the RETs
share and operational optimization. However, on the other side, we should also keep in mind that digital
technologies, which are likely to be functional to the presence of a
fully automated grid, generate risks for new types of addictions
\cite{takahashi2018behavioral}. Also, this represents a potential problem for individuals
privacy due to the growth of data and meta-data gathering, thus enabling
a dramatic, un-ethical level of individual insights through business
intelligence triangulations.\\
The misunderstanding of the philosophical premises and the presumable incomprehension of the consequences of the
digital tsunami -- from every perspective -- clearly generated by a
non-controlled market speculative pull, might tend to consolidate the
presence of a technology-based society, increasing the possibilities to
realize what could be tagged as `digital middle-age', that might consists
in an apparently advanced society characterized by contradiction where
the presence of vertical technological advances are in contraposition to
a compromised individuals wellbeing. To prevent this
not-desirable scenario, thus, to enable true societal progress, as
previously reported by other scientists \cite{bigerna2015overview}, we further underline
the imminent need of multidisciplinary cooperation increasing the
humanities and social sciences contribution to wisely address the smart
grid development, avoiding to let commercial speculation and un-ethical
choices to lead and shape our future society.

\bibliographystyle{unsrt}  
\bibliography{How_smart_is_the_grid}  



\end{document}